\documentstyle[12pt,epsf,captions]{article}
\def\singlespace 
{\smallskipamount=3.75pt plus1pt minus1pt
\medskipamount=7.5pt plus2pt minus2pt
\bigskipamount=15pa plus4pa minus4pt \normalbaselineskip=12pt plus0pt
minus0pt \normallineskip=1pt \normallineskiplimit=0pt \jot=3.75pt
{\def\smallskip {\vskip\smallskipamount}} {\def\medskip
{\vskip\medskipamount}} {\def\bigskip {\vskip\bigskipamount}}
{\setbox\strutbox=\hbox{\vrule height10.5pt depth4.5pt width 0pt}}
\parskip 7.5pt \normalbaselines} 

\def\middlespace
{\smallskipamount=5.625pt plus1.5pt minus1.5pt \medskipamount=11.25pt
plus3pt minus3pt \bigskipamount=22.5pt plus6pt minus6pt
\normalbaselineskip=22.5pt plus0pt minus0pt \normallineskip=1pt
\normallineskiplimit=0pt \jot=5.625pt {\def\smallskip
{\vskip\smallskipamount}} {\def\medskip {\vskip\medskipamount}}
{\def\bigskip {\vskip\bigskipamount}} {\setbox\strutbox=\hbox{\vrule
height15.75pt depth6.75pt width 0pt}} \parskip 11.25pt
\normalbaselines} 

\def\doublespace 
{\smallskipamount=7.5pt plus2pt minus2pt \medskipamount=15pt plus4pt
minus4pt \bigskipamount=30pt plus8pt minus8pt \normalbaselineskip=30pt
plus0pt minus0pt
\normallineskip=2pt \normallineskiplimit=0pt \jot=7.5pt
{\def\smallskip {\vskip\smallskipamount}} {\def\medskip
{\vskip\medskipamount}} {\def\bigskip {\vskip\bigskipamount}}
{\setbox\strutbox=\hbox{\vrule height21.0pt depth9.0pt width 0pt}}
\parskip 15.0pt \normalbaselines}

\newcommand{\be}{\begin{equation}}
\newcommand{\ee}{\end{equation}}
\newcommand{\bea}{\begin{eqnarray}}
\newcommand{\eea}{\end{eqnarray}}
\begin{document}
\begin{center}
\large {\bf Effects of 126 dimensional Higgs scalar on Bottom-Tau 
unification and quasi-infrared fixed point} \\ 
\vskip 1in
Biswajoy Brahmachari \\
\end{center}
\begin{center}
Department of Physics and Astronomy,\\
University of Maryland, College Park,\\
MD-20740, USA.
\\
\end{center}
\vskip 1in
{%\singlespace
\begin{center}
\underbar{Abstract} \\
\end{center}
{ {\small                                                                    
In the presence of ${\bf 126 + \overline{126}}$ Higgs multiplets in a 
SO(10) theory, the fermion masses get contributions from an induced 
vacuum expectation value (VEV) of a $SU(2)_L$ doublet residing in 
${\bf 126}$ which differentiates between quarks and leptons by a relative 
sign leading to a significant correction to the prediction of the mass
ratio of the bottom quark and the tau lepton for ranges of the mass of
this extra doublet. We perform a two-loop renormalization group analysis
of the minimal version of the one-step supersymmetric SO(10) model to
display this and re-calculate the corrections to the top quark mass in
the presence of such an induced VEV. We show that these effects make the
infra-red fixed point scenario consistent with experimental results. }
}
\newpage

\doublespace

In the minimal supersymmetric SO(10) grand unified theory (GUT), the 
fermions reside in the {\bf 16} dimensional spinorial multiplet while 
the low-energy doublet Higgs fields do in the {\bf 10} dimensional 
fundamental representation, and consequently in conventional wisdom, the 
fermion masses arise from the triliniar ${\bf 16 \times 16 \times 10}$ 
Yukawa coupling of the underlying GUT superpotential. This is a tightly 
constrained scenario which allows only one parameter $h_X$ for the third 
generation fermion masses in the Yukawa superpotential. At low energy the 
Yukawa couplings get split by renormalization group effects of the mass-less 
fields below the GUT scale, while the prediction of the mass ratio of the 
bottom quark and the tau lepton and its comparison to the experimentally 
measured values at their own scales has been extensively studied in the
literature \cite{lang}.

On the other hand, it has to be emphasized that at present the top quark 
mass is determined \cite{topmass} with in an error of approximately  $\pm 
12$ GeV only, where as, the present central value (173 GeV) has also reduced 
from earlier \cite{topold} analysis (176 GeV for CDF and 199 GeV for {$\rm 
D\O$ collaborations}) apparently constraining the top quark Yukawa 
coupling at the GUT scale considerably. In fact, the minimal version of 
SO(10) model described above when $h_X$ is at the infra-red fixed point 
\cite{fixt} region, predicts too large a value for the top quark mass 
$m_t^{pole}(m_t)\sim 200$ GeV. A note on the prediction of the top quark 
mass in the infrared fixed point scenario vis-a-vis the current 
experimental range of the top quark mass with the variation of $\tan 
\beta$ can be found in Ref\cite{mpl}. We have summarized these results in 
Figure (1.A) in the case of minimal SO(10) GUT 
with the variation of $h_t=h_b=h_\tau \equiv h_X$ at the unification 
scale. In such a scenario $\tan\beta$ is approximately ${m_t \over m_b} 
\sim 60$. It is, thus, of interest to re-evaluate the fixed point 
scenario in the SO(10) GUT to make sure whether any natural contribution 
to the quark masses even in the minimal version\footnote{ A more minimal 
choice will be to use a pair of $16$ Higgs instead of $126$ to break the 
right-handed symmetry. However this will also break R-parity spontaneously 
and neutrino-masses will arize from non-renormalizable operators. We are not 
considering this option.}with a single $10$ 
and a pair of $126$ has been overlooked which can make the predictions 
consistent with $m_b$ and $m_t$ and lower values of $\tan\beta$ of the 
order of unity. 

\begin{figure}
\begin{tabular}{cc}
\epsfysize=8.5cm \epsfxsize=8.5cm \hfil \epsfbox{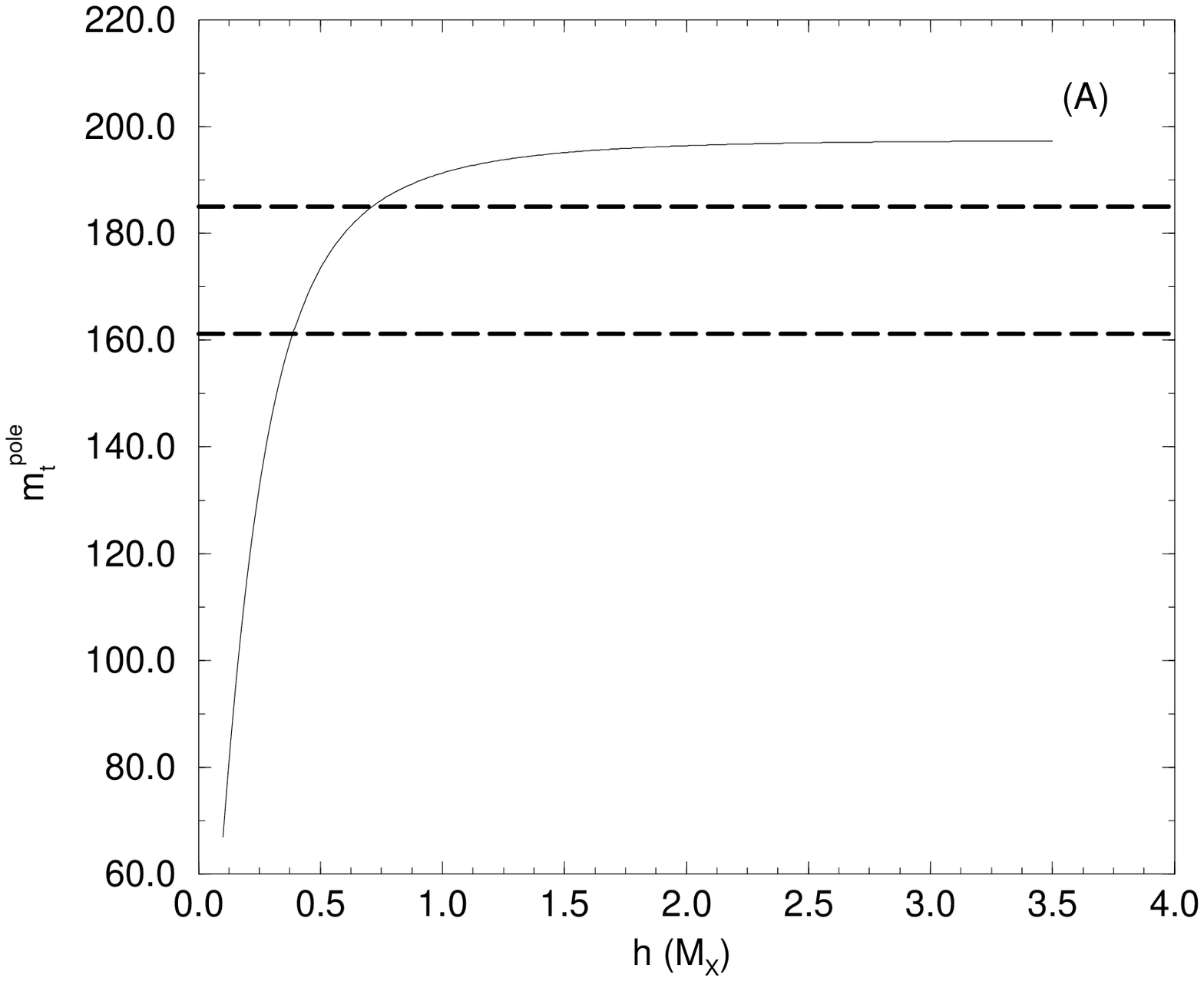} \hfil
& \epsfysize=8.5cm \epsfxsize=8.5cm \hfil \epsfbox{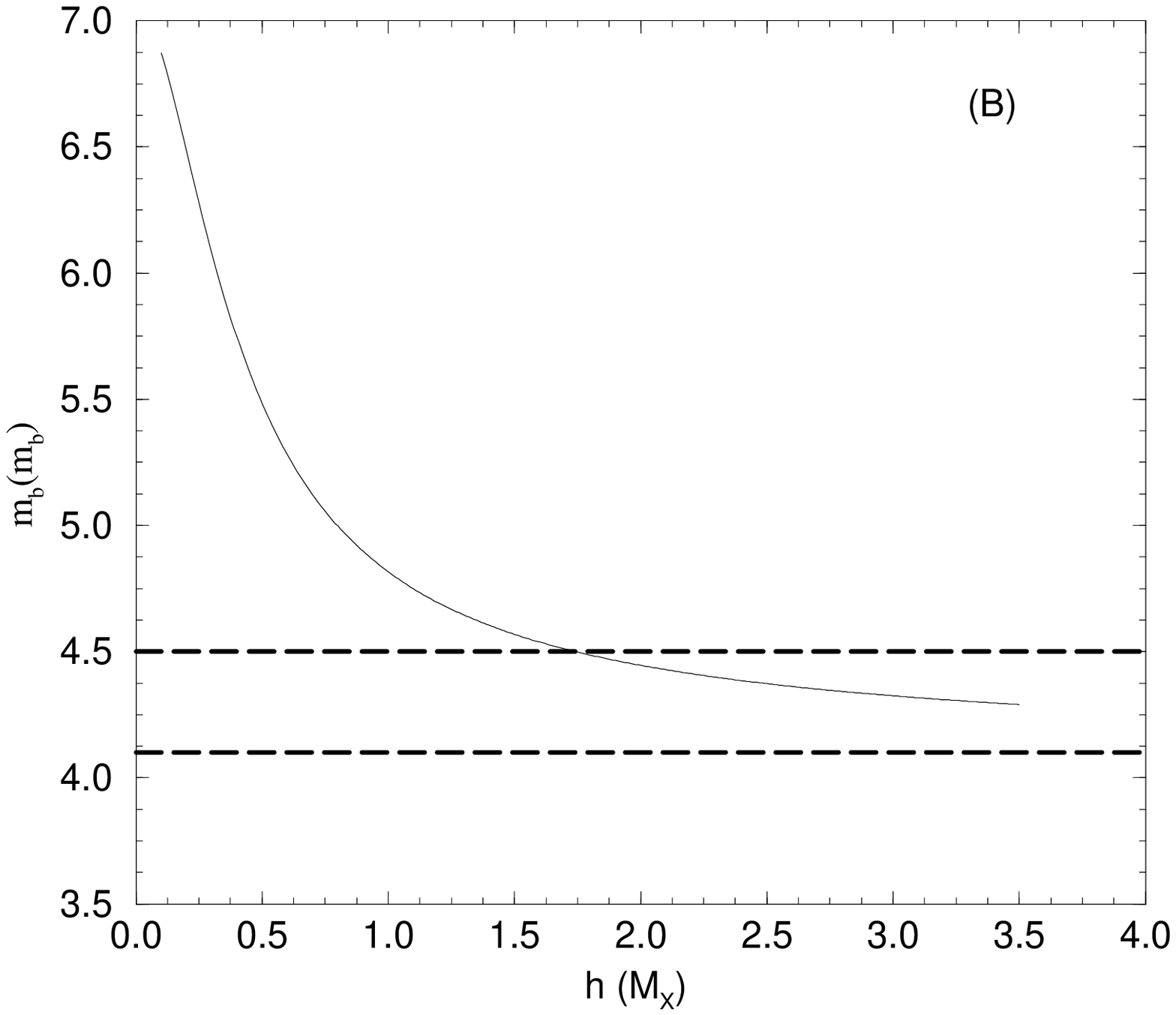} \hfil\\
\end{tabular}
\caption{(A) The variation of the top quark (B) bottom quark masses with
respect to the unified GUT scale Yukawa coupling in SO(10) model with
minimal Higgs choice excluding the effects of $\overline{126}$. The fixed
point scenario is at the upper right corner of the figure. The prediction
supersedes experimental measurement for $h_X < 0.53$ or $Y_X \equiv 
{h^2_X \over 4 \pi} < 0.022$. It is notable that in the minimal SO(10) 
model, complete Yukawa unification forces $\tan \beta \sim {m_t \over 
m_b} \sim 60$. While quoting $m_b$ and $m_t$ we have used the same value 
of $M_\Sigma$.} \end{figure}

In an SO(10) GUT, the intermediate gauge symmetry \cite{int}
is most naturally broken by a $SU(2)_R$ Higgs fields ${\bf 3 + 
\overline{3}}$ which are parts of the SO(10) multiplets ${\bf 
126+\overline{126}}$. According to the Appelquist-Carazzone decoupling 
theorem \cite{apcor}, the triplet parts of ${\bf 126}$ get mass at the 
breaking scale of $SU(2)_R$, we call it $M_I$, while the rest of the 
fields get mass at the next higher scale, which is the unification scale 
$M_X$ in our notation. Among the various components of ${\bf 126}$, we 
need to note the existence of a left handed doublet $\Sigma$ apart from
the usual doublet H for our purposes. In is easy to see an interesting 
property of this $SU(2)_L$ doublet component of ${\bf 126}$; if we compare 
it's expansion in terms of the representations of the SO(10) subgroup 
$SU(2)_L \times SU(2)_R \times SU(4)_C$ to the expansion  of {\bf 10}, as,
\begin{eqnarray} 126 & \longrightarrow & (1,1,6) + (1,3,10) + (3,1,10) + 
{\bf (2,2,15)}, 
\nonumber\\ 10  & \longrightarrow & (1,1,6) + {\bf (2,2,1)}. 
\nonumber 
\end{eqnarray}
As $\Sigma$ forms an adjoint of $SU(4)_c$, and as it's VEV has to be 
$SU(3)_c$ preserving, the VEV should be of the form,
{ \small
\begin{equation}
\langle \Sigma \rangle = \Sigma_0 \pmatrix{1&&& \cr&1&&\cr&&1&\cr&&&-3},
\end{equation}
} 
which differentiates between the quarks and leptons by a factor of three 
as well as a relative sign to maintain tracelessness. Hence, unlike 
the VEV of H, a non-zero VEV of $\Sigma$ affect the masses of quarks and 
leptons in mutually opposite directions \cite{babu}, and hence, in 
particular has interesting effects on the mass ratio of $m_b/m_\tau$. In 
this letter we study this possibility. 

Let us consider the part of the SO(10) invariant superpotential,
\begin{equation}
{\cal W} = h~16_F~16_F~10_H + f~16_F~16_F~{\overline{126}}_H 
+ \lambda_1~126_H~\overline{126}_H~210_H + 
\lambda_2~126_H~10_H~210_H + ... \nonumber
\end{equation}
For simplicity we take all the couplings to be real. The F term with 
respect to ${\bf 
210}$ gives the quartic Higgs coupling in the SO(10) scalar potential 
$\lambda_1\lambda_2~126~\overline{126}~126~10 \equiv \lambda ~126 
~\overline{126} ~126 ~10$. When {\bf 126} gets a VEV ($v_R$) 
of the order of $M_I$ and {\bf 10} gets a VEV of the order $m_Z$ the fermions
get a correction to their mass through the second term in ${\cal W}$ written 
in terms of the mass of the doublet in ${\bf \overline{126}}$, 
$M_\Sigma$, as, 
\begin{equation}
\Sigma_0 = \lambda_1 \lambda_2 {v^2_R \over M^2_\Sigma} ~\langle H \rangle 
= \lambda {M^2_I \over \alpha^2_R M^2_\Sigma}~\langle H \rangle.  
\end{equation}

We will consider minimal SO(10) model. It is well-known that the minimal
model leads to an one-step unification \cite{amaldi} with the 
unification scale at $M_X = M_I \sim 10^{16.4}$ GeV, and 
$\alpha_L(M_X)=\alpha_R(M_X)=\alpha_G=0.04306$ in a two-loop evolution 
of the gauge couplings starting from the well measured values 
$\alpha_{L}(m_Z)=0.03371$ and $\alpha_L(m_Z)=0.01696$, with a prediction of 
$\alpha_s(m_Z)= 0.125-0.126$ requiring gauge coupling unification 
\cite{bagger}. Using these we re-express \footnote{From now on we 
denote $\lambda~f\equiv f$.} the VEVs of $\Sigma$ ($\kappa$) in terms of 
those of H (v) as, 
\begin{equation}
\kappa_{1,2} = {M^2_X \over \alpha^2_G M^2_\Sigma}~v_{2,1} 
= f~539.17 ~{M^2_X \over M^2_\Sigma}~v_{2,1}. \label{reln}
\end{equation}

Where the Higgs fields with subscript 2 couples to the top quark
and those with subscript 1 couples to the bottom quark. Note that 
hypercharge balance at the four-point vertex 
$f~\Sigma~\Delta~\overline{\Delta}~H$ requires that $v_2$ induces 
$\kappa_1$ and vice-versa. And also, when supersymmetry in unbroken, 
neither (2,2,1) of 10 nor (2,2,15) of $\overline{126}$ has VEVs. Now we are 
led to the relation for the ratio $m_b \over m_\tau$, \begin{equation}
{m_b(m_b) \over m_\tau(m_\tau)} = \eta_b~[{h_b ~v_1 + f_b ~\kappa_1 \over 
h_\tau ~v_1 -3 ~f_\tau ~\kappa_1}] = \eta_b~[{h_b + 539.17 ~{M^2_X \over 
M^2_\Sigma}~ f_b \over h_\tau -1617.51~{M^2_X \over M^2_\Sigma}~ f_\tau}],  
\label{mbpred}
\end{equation}
while for the top quark mass we get,
\begin{equation}
m_t({m_t})=[h_t + 539.17 ~{M^2_X \over M^2_\Sigma}~ f_t ]~ v_2
~~~{\rm and}~~~m^{pole}_t=m_t(m_t)~[1 + {4 \over 3 \pi} \alpha_s]. 
\label{mtpred} 
\end{equation}
In Eqn. (\ref{mbpred}) and (\ref{mtpred}) we have redefined
\begin{equation}
f_b=f_\tau=f~\tan \beta~~;~~f_t=f~\cot\beta~~;~~\tan\beta={v_2 \over 
v_1}. 
\end{equation}
Note that the familiar experimental quantity $\tan \beta$ gets absorbed in
the re-definition of the coupling $f$. In the mass ratio of $m_b$ to 
$m_\tau$ the conventional term $h$ can only dominate if the ratio, 
\begin{equation}
h_\tau >> 1617.51~{M^2_X \over M^2_\Sigma}~ f_\tau~~~{\rm or}~~~{M^2_X 
\over M^2_\Sigma} ~<< {h_\tau \over f_\tau} ~0.625 \times 10^{-3}.
\end{equation}
Whenever $M_\Sigma < M_X$ the ratio ${h_b \over h_\tau}$ stays close to
$-{1 \over 3}$ which is far from experimental value. This implys that
the mass scale of $\Sigma$ should be larger than $M_X$. Consequently the 
couplings $f$ do not renormalize below the scale $M_X$ staying constant 
up-to the low energy scale. While extrapolating the couplings $h_t$, 
$h_b$ and $h_\tau$ at the two-loop order \cite{jones} we simply have to 
use the beta function coefficients of the gauge and Yukawa couplings of 
MSSM from the scale $M_X$ to the scale $m_t$.

We note that the VEVs still cancel among the numerator and the 
denominator of the ratio of the bottom and the tau Yukawa couplings.
Secondly, in the limit when the second term dominates the $b-\tau$ mass 
ratio, the ratio becomes $-{1 \over 3}~\eta_b$ where as when the first term 
dominates the ratio we recover (for $\alpha_s=0.125$) the known value about 
1.6. There is a jump or a monotonous variation of the ratio depending on 
the sign of $f$ which is the parameter space of our interest. 

Now let us turn to the RGE analysis. The system of Yukawa evolution 
equations is a coupled one. If we start from an initial value of the 
Yukawa couplings at the GUT scale $Y(M_X)={h^2(M_X) \over 4 \pi}$, and 
evolve the couplings to the electroweak scale using Renormalization Group 
Equations (RGE), the pattern of dependence is as follows. The Yukawa part 
of the beta functions tend to decrease the value $Y(m_t)$ where as the 
influence of the gauge couplings, especially the QCD coupling is to 
increase the value of $Y(m_t)$. Consequently, if we start from a 
non-perturbative (large) value of the Yukawa couplings at the GUT scale, 
numerically speaking, the Yukawa effects dominate the beta-functions at 
higher scales making its value at $m_t$ smaller than the value obtained 
if we start from a relatively small value of $Y(M_X)$. This property of the 
Yukawa evolution equations has a consequence for the prediction of the 
b-quark mass. If the unified $b-\tau$ Yukawa coupling is small at the 
GUT scale, \footnote{In the case of SU(5) GUT where the unified $b-\tau$ 
coupling will come from a separate term in the SU(5) superpotential from 
that the top quark Yukawa term} the prediction of the b-quark mass 
becomes more than even the liberal experimental upper-bound \cite{lang} of 
about 5.2 GeV whereas, in the fixed point scenario, these `bare' 
predictions from RGE of the bottom quark mass emerges in the correct 
range; which has been summarized in Figure (1.B). The `bare' predictions 
get `dressed' up by loop corrections dominated by graphs involving gluinos 
[See Eqn. (\ref{mbcorr})].

There have been various attempts to modify these prediction. Using the 
renormalization effects of the Majorana Yukawa coupling \cite{brignole} 
pushes the ratio in the reverse direction and consequently the b-quark 
mass increases further. One can consider
the renormalization effects of adjoint Higgs scalars at an intermediate 
scale \cite{marstring}, which are the so called remnants of string 
theory, or the effects of an intermediate scale breaking of a gauge 
symmetry like the left-right symmetry \cite{rabiplb} and the related 
renormalization effects on the ratio.

We have plotted the results of our numerical analysis using the formulas 
given in Eqn. (\ref{mbpred}) and Eqn. (\ref{mtpred}) in Figures (2) and
(3) for positive and negative values of the couplings $f_t$, $f_b$ and 
$f_\tau$. Assuming the mass $m_\tau(m_\tau)=1.777$ we can calculate the 
value of $m_b(m_b)$ which has been plotted in figure (2.A) and (3.A). In 
doing so, we have extrapolated the mass of the b-quark from the scale 
$m_t$ to its own scale by using the factor $\eta_b$ defined in reference 
\cite{berg} which includes 3-loop QCD corrections and one loop QED 
corrections. The top quark mass predictions have been plotted in Figure 
(2.C) and (3.C). 

Concerning the top quark mass prediction, we see the dotted horizontal lines 
( f=0; h at the fixed point ) predict $m_t \sim 200~$ GeV  which is outside
the present ${\rm D\O}$ range \cite{topmass} of $173 \pm 5.6~({\rm stat}) 
\pm 6.2~({\rm syst})$ GeV. Inclusion of the effects of the coupling f 
[Figure (3)] reduces the top mass (when f is negative) in the experimentally 
allowed range. The prediction of the b-quark mass also reduces 
accordingly. However, we note that there is a large uncertainty 
\cite{sarid} in the prediction of the  bottom  quark mass due to 
one-loop gluino graphs; correcting the prediction to $m_b= h_b~174 \cos 
\beta + \delta m_b$, where, 
\begin{equation}
{\delta m_b \over m_b}= {8 \over 3} \alpha_s {\tan \beta \over 4 \pi}
{m_{\tilde{g}} \mu \over m^2_{eff}}, \label{mbcorr}
\end{equation} 
and where $m_{eff}$ is the mass of the heaviest superpartner in the loop. 
The ratio in the LHS can be of order unity, consequently, b-quark mass 
cannot be predicted well unless the superpartner spectrum is fixed. Below 
we have estimated the magnitude of the correction for a simple 
supersymmetric spectrum. The variation in $\tan \beta$, which is absorbed 
in the coupling $f$ shows up here in the supersymmetric 
correction to $m_b$.

Figure (3) summarizes the results of our interest. In the limit $f=0$, 
denoted by bold dashed lines, the prediction of $m_b$ is within the 
experimental bounds but that of $m_t$ is too high. Small $f_b$ still 
gives good values $m_b$, however a moderate value of $f_t$ improves the 
prediction of $m_t$. For example consider $f_b=0.1$ and $f_t=0.4$. From 
Figure (3) we get that for $M_\Sigma=10^{17.75}$ we get both $m_b$ and 
$m_t$ in the experimentally allowed range. The 
parameter $\tan \beta$ is given by, 
\begin{equation}
\tan^2 \beta = {f_b \over f_t}. \label{tanb}
\end{equation}
We note that for the case of large $\tan \beta$, ie, $f_b >> f_t$ we 
would have a high value of $m_t$ and low value of $m_b$. And 
of-course in the limit $M_\Sigma \rightarrow \infty$ we will recover the 
prediction of the minimal SO(10) prediction at the right end of the graphs.
However once we start to lower $M_\Sigma$ the parameter space of $f_b << 
f_t$ or equivalently low $\tan \beta$ gives good fit to top and bottom 
quark masses. Once $\tan \beta$ is fixed from Eqn.(\ref{tanb}) we can 
calculate the correction to $m_b$. In the case given above $\tan 
\beta=1/2$ and hence for a simple case where the $\mu$ parameter is equal 
to a degenerate susy spectrum the correction term is $\delta m_b = 0.054 
(0.048)$ GeV when $\alpha_s=0.125(0.112)$. When $\tan \beta=4$ the 
correction is $\delta m_b=0.435(0.389)$ GeV and in the case of $\tan 
\beta=10$ the correction is $\delta m_b=1.08 (0.974)$. Thus we see that 
for values of $\tan \beta=O(1)$ the effect of the Majorana Yukawa 
coupling $f$ to the masses of the quarks can reduce the prediction of the 
top quark pole mass bringing it within $173 \pm 6$ GeV, while still 
keeping the prediction of $m_b$ within experimental limits. This is the 
main result of the paper.

The left handed neutrino mass is inversely proportional to the mass of 
the right handed triplet Higgs breaking the right handed SU(2) symmetry 
and approximately proportional to the square of the up quark mass of the 
given generation via see-saw mechanism. We have calculated 
$m_{\nu_\tau}$. For $f$ in the range of $10^{-1}$ the mass of the tau 
neutrino is of the order of $10^{-6}$ GeV. Note that it is a simple 
analysis taking the third generation couplings only, whereas in the
neutrino sector it is better to calculate the eigenvalues of the full
$3\times3$ mass matrices. 

At two-loop level, the unification of couplings occur for 
$\alpha_s(m_Z)=0.125$. In this study we have not considered threshold 
effects on the RGE running due to the spread of supersymmetric spectrum 
or the super heavy spectrum at the GUT scale. The effect of the susy 
spectrum is always to increase the prediction of $\alpha_s(m_Z)$ whereas 
the GUT scale threshold effects can either increase or decrease the 
prediction of the strong coupling. The world average of $\alpha_s (m_Z) 
\sim 0.118$. Our value of 0.125 is barely consistent with the world 
average including the error-bars. Had we included the GUT scale threshold 
corrections in the running of the gauge couplings\cite{bagger}, we would 
have gotten a lower value of $\alpha_s$. For the purpose of $b-\tau$ 
unification this would lead to a slightly lower value of $m_b$ as the QCD 
renormalization increases $h_b$ at low energy. However, a suitable 
variation in $f_b$ would compensate it back to the values shown 
in Figures (2) and (3). The same effects of the `Majorana' coupling 
hold for the $\alpha_s$ dependence of prediction of $m_t$. 

Before we conclude, we would like to make the following observation. In 
the minimal SU(5) the top quark Yukawa is not required to unify with the 
bottom quark Yukawa coupling at the unification scale because they 
are embedded in different multiplets of SU(5); however the bottom 
quark and the tau lepton are embedded in the same multiplet of 
SU(5) and they must be unified. Once the top quark Yukawa coupling is 
kept at the infrared fixed point, and the magnitude of the unified 
bottom-tau Yukawa coupling is varied (which is equivalent to varying 
$\tan \beta$) there are interesting effects on the prediction of $m_b$ due 
to the variation in self renormalization of the bottom-tau Yukawa 
coupling. In fact Yukawa unification occurs at specific regions of the 
$\tan\beta- m_b$ plane in the bottom $\rightarrow$ up approach. On the 
other hand in the minimal SO(10) Yukawa unification scenarios, we always 
have large $\tan\beta \sim {m_t \over m_b} \sim 60$. In our case the mass 
of the quarks and leptons arise due to combined contributions from the 
VEVs of the $10$ Higgs and the $126$ Higgs fields. Hence given an 
experimental value of $m_b$, in this SO(10) scenario, $complete$ Yukawa 
unification in the $10$ and $126$ sectors can occur independent of $\tan 
\beta$. 

In conclusion, in a SO(10) theory there exists the gauge invariant
coupling of the ${\bf \overline{126}}$ with the fermions. As the doublet
scalar residing in ${\bf \overline{126}}$ recieves an induced VEV  
when the triplet in the same representation gets a VEV at the scale
$M_R$, it is important to study it's consequences to $b-\tau$ unification of 
Yukawa couplings. Unlike the normal Yukawa couplings $h$ the coupling to 
${\bf \overline{126}}$ stays flat below the unification scale and so, for 
large ranges of values of the parameter $M_\Sigma$, the mass of the 
$SU(2)_L$ doublet part of ${\bf \overline{126}}$, the effect of the 
coupling $f$ changes the ratio ${m_b \over m_\tau}$ which to the best 
of our knowledge was un-noticed in the literature. We have summarized 
these results in Figures (2.A) and (3.A). The top quark mass [Figure (3.C)] 
reduces to the experimentally acceptable range or increases further 
[Figure (2.C)] away from experimental values depending on whether we have 
negative or positive value of the coupling $f$. 

\noindent It is a pleasure to acknowledge the hospitality of Saha 
Institute of Nuclear Physics, Calcutta, India for a two-month visit; 
discussions with R. N. Mohapatra and private communications with K. S. 
Babu. This work has been supported by a grant from the National Science 
Foundation. 

\begin{figure}
\begin{tabular}{cc}
\epsfysize=9cm \epsfxsize=9cm \hfil \epsfbox{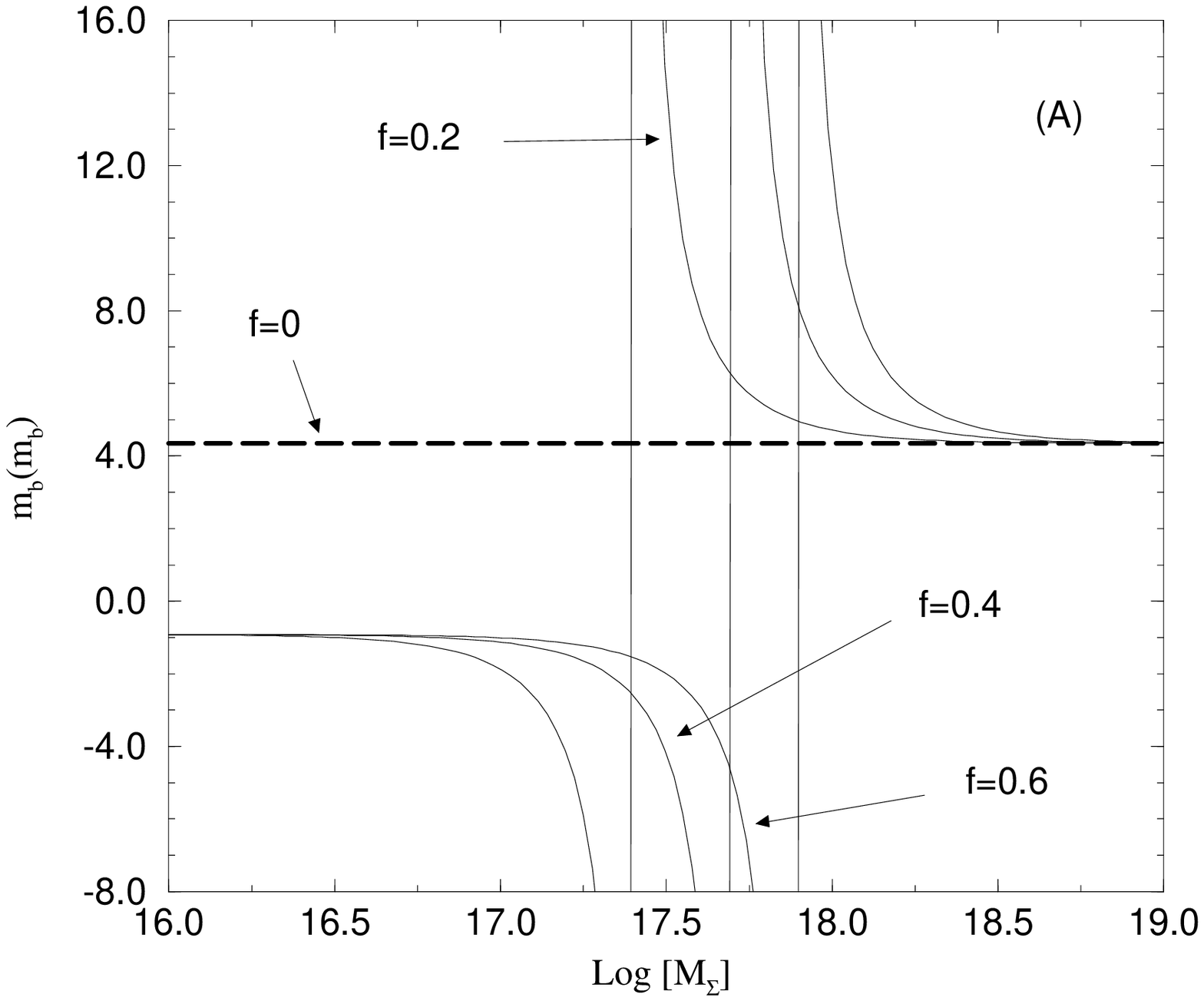} \hfil
& \epsfysize=9cm \epsfxsize=9cm \hfil \epsfbox{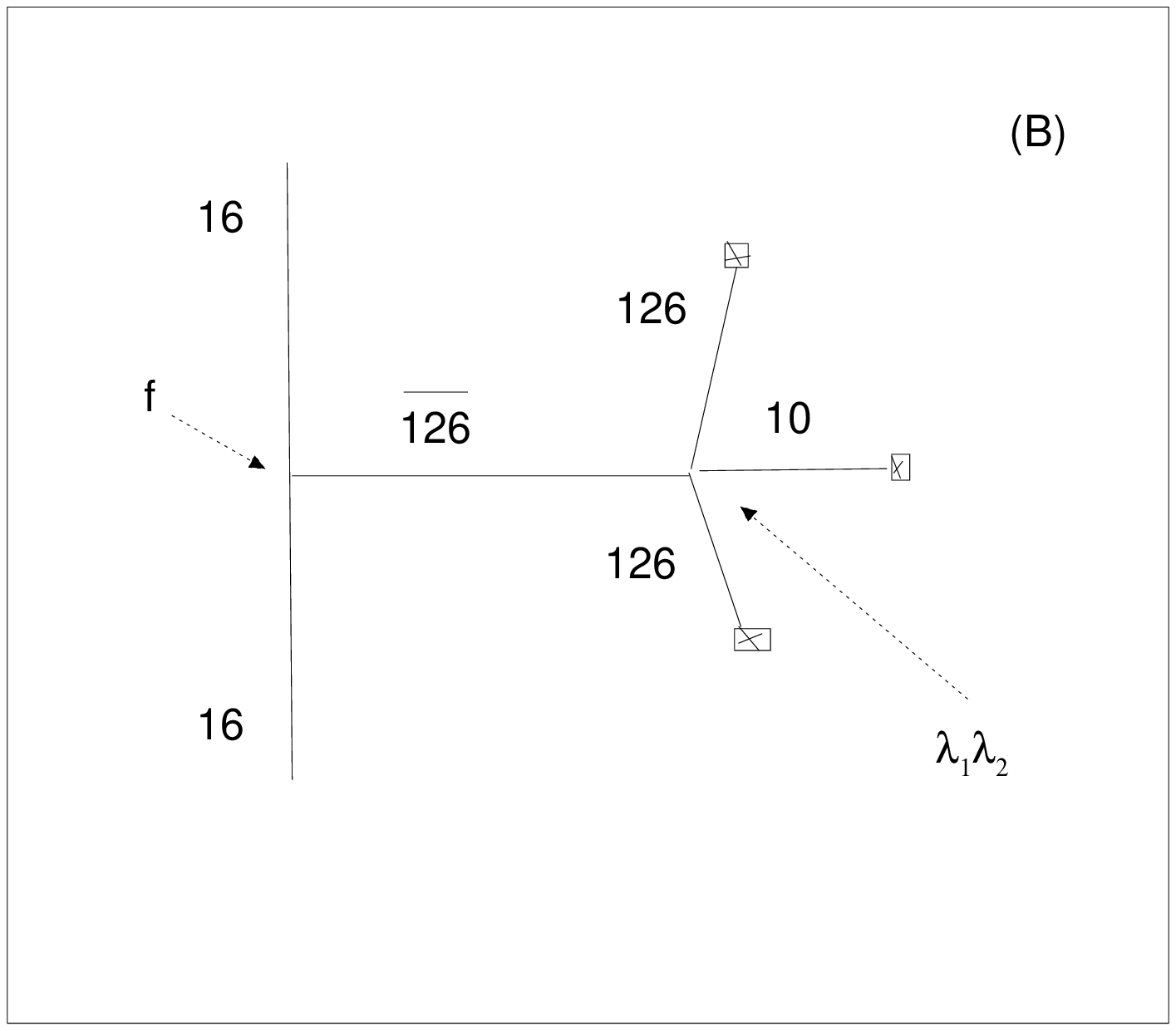} \hfil\\
\epsfysize=9cm \epsfxsize=9cm \hfil \epsfbox{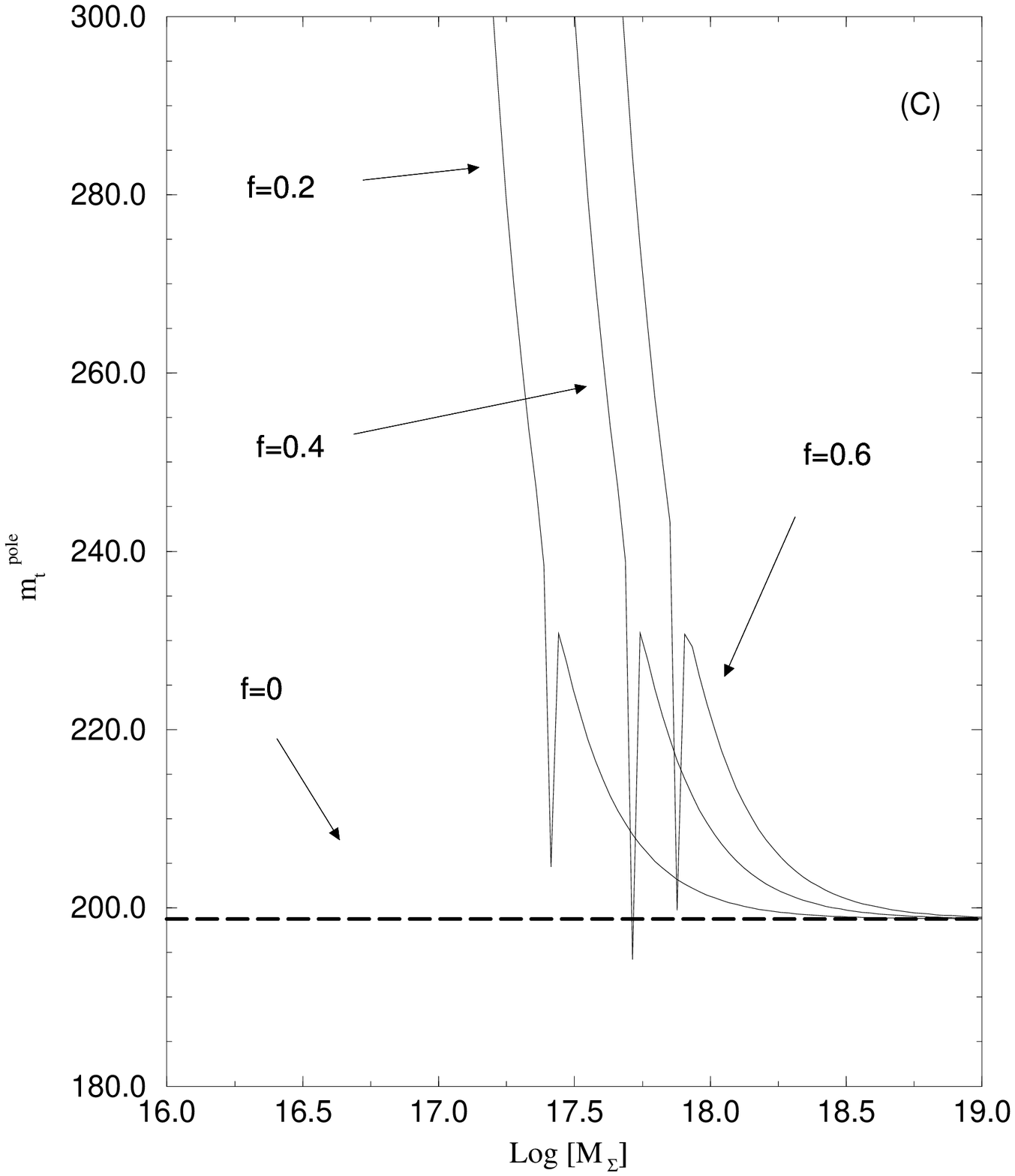} \hfil
& \epsfysize=9cm \epsfxsize=9cm \hfil \epsfbox{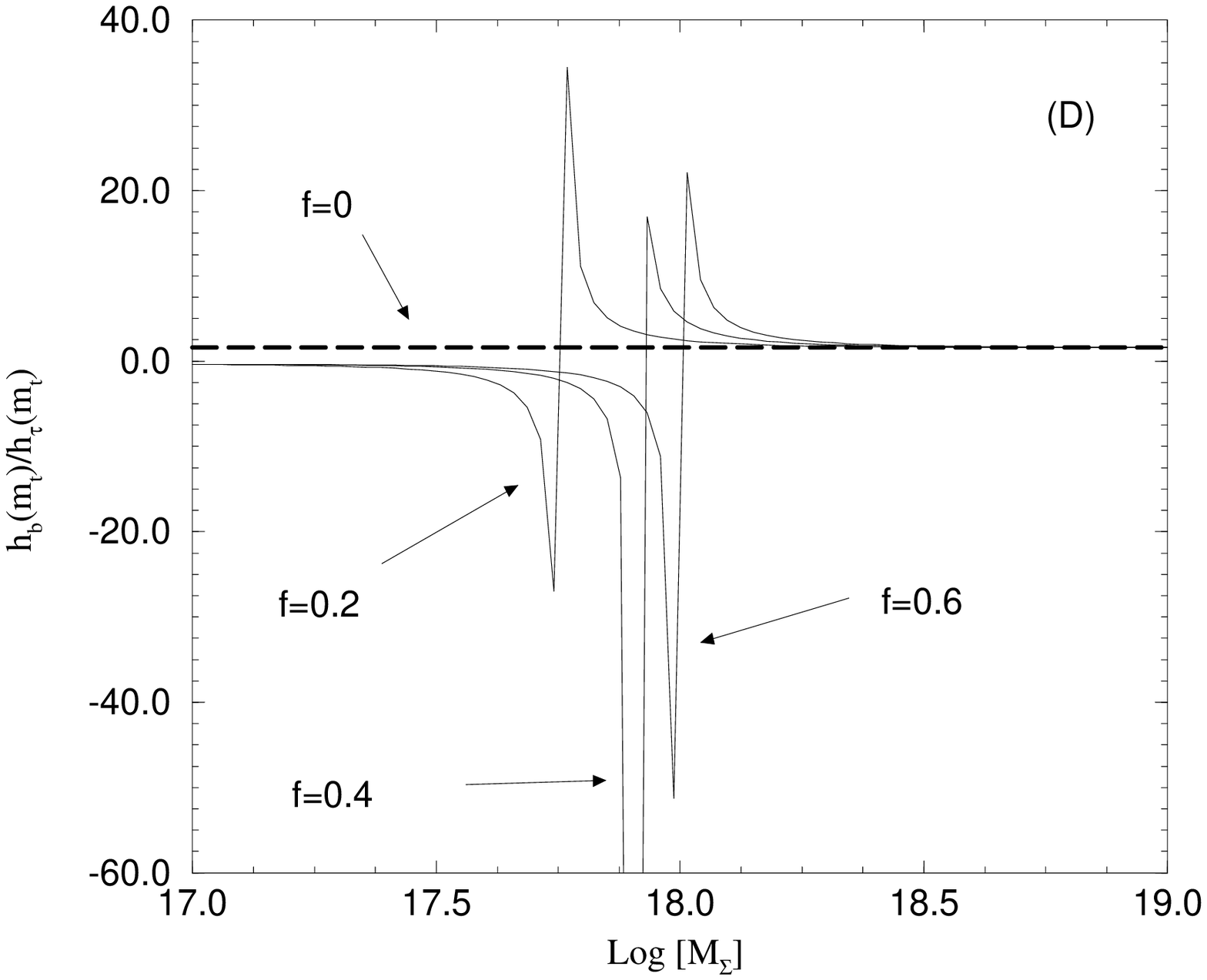} \hfil
\end{tabular}
\caption{(A) The prediction of b-quark mass as a function of the mass of
the extra doublets $M_\Sigma$ when $ f \equiv f_b \equiv f_\tau$ is 
positive (B) the Feynman graph which leads to the induced mass of the 
fermions (C) The prediction of the top-quark pole mass where $f \equiv 
f_t$(D) the prediction of the ratio of the b-quark and $\tau$-lepton 
mass. Figure (A) can be recovered from (D) assuming $m_\tau=1.777$ GeV. 
The dashed lines are corresponding MSSM values.} 
\end{figure} 

\begin{figure}
\begin{tabular}{cc}
\epsfysize=9cm \epsfxsize=9cm \hfil \epsfbox{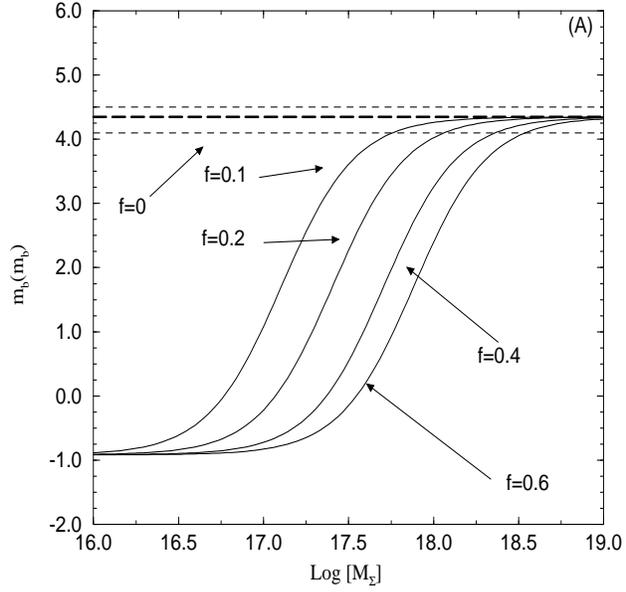} \hfil
& \\
\epsfysize=9cm \epsfxsize=9cm \hfil \epsfbox{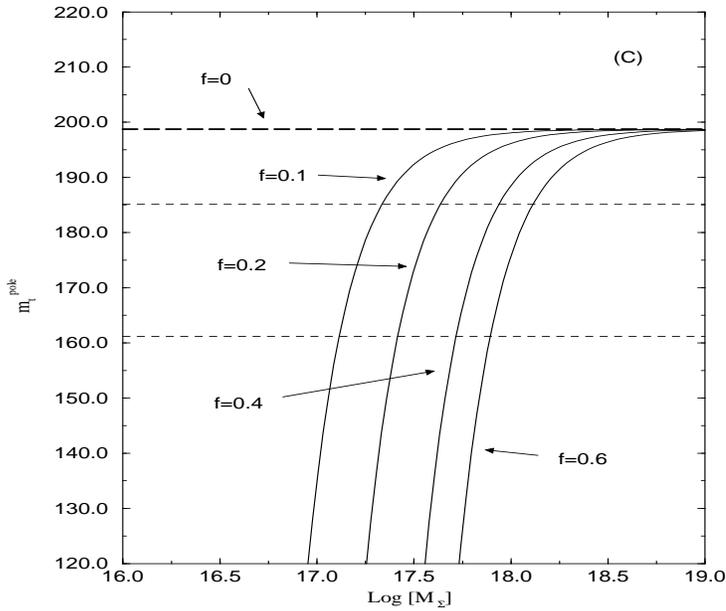} \hfil
& \epsfysize=9cm \epsfxsize=9cm \hfil \epsfbox{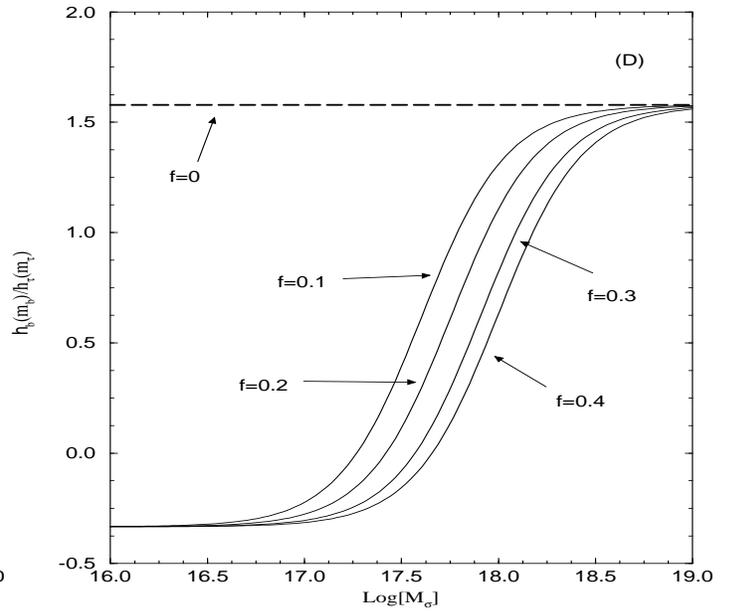} \hfil
\end{tabular}
\caption{Same as Figure (1) but the effect of the $\overline{126}$ is
in the reverse direction reducing the top-quark mass prediction. The
b-quark mass prediction is also reduced, however remains within 
experimental limits modulo theoretical uncertainties.} 
\end{figure}

\end{document}